\documentclass{article}
\usepackage{amsmath,graphicx,mlspconf,mathtools,bm}
\usepackage{algorithm,algorithmic}
\usepackage{amssymb,setspace}
\newcommand{\ri}{\ensuremath{\mathrm{i}}}
\newcommand{\E}{\ensuremath{\operatorname{E}}}
\usepackage{etoolbox}
\apptocmd{\thebibliography}{\setlength{\itemsep}{-2.89pt}}{}{}

\title{EXACT SIMULATION OF NONCIRCULAR OR IMPROPER COMPLEX-VALUED STATIONARY GAUSSIAN PROCESSES USING CIRCULANT EMBEDDING}
\twoauthors
  {Adam M. Sykulski\thanks{The work of A. M. Sykulski was supported by a Marie Curie International Outgoing Fellowship within the 7th European Community Framework Programme.}}
	{Department of Statistical Science,\\University College London, UK}
  {Donald B. Percival}
	{Applied Physics Laboratory,\\University of Washington, Seattle, USA}

\begin{document}
%

\maketitle
\begin{abstract}
This paper provides an algorithm for simulating improper (or noncircular) complex-valued stationary Gaussian processes. The technique utilizes recently developed methods for multivariate Gaussian processes from the circulant embedding literature. The method can be performed in $\mathcal{O}(n\log_2 n)$ operations, where $n$ is the length of the desired sequence. The method is exact, except when eigenvalues of prescribed circulant matrices are negative. We evaluate the performance of the algorithm empirically, and provide a practical example where the method is guaranteed to be exact for all $n$, with an improper fractional Gaussian noise process.
\end{abstract}
\begin{keywords}
Circulant embedding, improper, noncircular, complex-valued, fractional Gaussian noise\\
\end{keywords}
\noindent
\copyright 2016 IEEE. Personal use of this material is permitted. Permission from IEEE must be obtained for all other uses, in any current or future media, including reprinting/republishing this material for advertising or promotional purposes, creating new collective works, for resale or redistribution to servers or lists, or reuse of any copyrighted component of this work in other works.
\section{Introduction}
\label{sec:intro}
Bivariate time series are commonly represented as complex-valued signals in numerous applications including wireless communication \cite{martin2004complex}, meteorology and oceanography \cite{gonella1972rotary}, and functional Magnetic Resonance Imaging (fMRI) \cite{rowe2005modeling}. There have been important methodological developments in machine learning and signal processing for complex-valued signals, see e.g. the books by \cite{mandic2009complex,schreier2010statistical}. There is a growing interest in analyzing complex-valued processes that are said to be {\em improper} (see e.g. \cite{picinbono1997second,schreier2003second,rubin2008kinematics}), as opposed to {\em proper}, concepts which we define formally in this paper. Improper processes are sometimes referred to as noncircular processes, and proper processes as circular processes \cite{mandic2009complex,ollila2008circularity}. A proper process can be interpreted as a second order process whose statistics are rotationally invariant when viewed in the complex plane, whereas improper processes are not, and the latter are seen to be more realistic models for applications \cite[Section 1.9]{schreier2010statistical}. Numerous improper complex-valued stochastic models have been proposed, e.g. \cite{picinbono1997second,navarro2009estimation,mohammadi2015improper,tobar2015modelling,sykulski2015improper}, for applications including climate and seismology, and in this paper we provide an algorithm for their efficient simulation.

For efficiently simulating Gaussian processes, the technique of circulant embedding \cite{davies1987tests,wood1994simulation} is receiving increasing attention. The underlying concept is that the covariance matrix of any regularly sampled stationary process is Toeplitz, and hence can be embedded into a larger matrix that is {\em circulant}. A circulant matrix has the useful property that the matrix can be diagonalized via Fourier Transforms, such that Gaussian sequences can be simulated in $\mathcal{O}(n\log n)$ operations, where $n$ is the length of the desired sequence. This is faster than the alternative method of Cholesky factorization which requires $\mathcal{O}(n^2)$ operations. The circulant embedding technique is exact, in that the simulated sequence has statistical properties that exactly match the model, there are however instances where exactness is not guaranteed, which has been the subject of further investigation, see e.g. \cite{dietrich1997fast,craigmile2003simulating}.

The method of circulant embedding was extended to simulating complex-valued processes in \cite{percival2006exact}. This procedure is only applicable to proper processes however. In this paper we extend the method to improper processes by incorporating techniques developed in \cite{helgason2011fast} for multivariate processes. This connection was also recently mentioned in \cite{coeurjolly2016fast}, although this paper primarily focuses on revisiting the problem of simulating proper processes. The contribution of our paper is to provide full details of the $\mathcal{O}(n\log_2 n)$ generating algorithm for improper processes, together with a practical example, and an empirical evaluation of its performance.

For improper processes modeled in the {\em frequency} domain, an exact simulation algorithm is provided in \cite{chandna2013simulation}. Here in this paper we detail an exact simulation algorithm for improper processes specified in the {\em time} domain by their covariance structure. The distinction is important, because processes specified analytically in one domain cannot in general be expressed analytically in the other. Therefore converting from one domain to the other often requires an approximation, thus losing the exactness of the procedure (see also \cite[Section 3.4]{helgason2011fast}). We also note approximate techniques for simulating improper processes in \cite{rubin2007simulation} using widely-linear filters \cite{picinbono1995widely}. Circulant embedding, on the other hand, can be used to exactly generate finite sequences from infinite-order non-Markovian Gaussian processes, as we shall demonstrate in this paper.

\section{Preliminaries}
\label{sec:preliminaries}
Consider a complex-valued stationary Gaussian process defined by $z_t=x_t+\ri y_t$, where $t\in\mathbb{Z}$. We assume without loss of generality, that $z_t$ is zero mean, as a non-zero
mean can be added to the simulated sequence {\em a posteriori}.

The process $z_t$ is fully specified by the autocovariances and cross covariances of $x_t$ and $y_t$ respectively, given by
\begin{align}
s_{xx}(\tau) &= \E\{x_{t+\tau} x_t\} \nonumber \\
s_{yy}(\tau) &= \E\{y_{t+\tau} y_t\} \nonumber \\
s_{xy}(\tau) &= \E\{x_{t+\tau} y_t\}
\label{eq:bivariateacvs}
\end{align}
for all $\tau\in\mathbb{Z}$, where $\E\{\cdot\}$ is the expectation operator. We note that $s_{xx}(\tau)=s_{xx}(-\tau)$ and $s_{yy}(\tau)=s_{yy}(-\tau)$, but in general $s_{xy}(\tau)\neq s_{xy}(-\tau)$. Equivalently, the process $z_t$ can be fully specified by its own autocovariance and {\em complementary} covariance sequence, specified respectively by
\begin{align}
s_{zz}(\tau) &= \E\{z_{t+\tau} z^\ast_t\} \nonumber \\
r_{zz}(\tau) &= \E\{z_{t+\tau} z_t\}
\label{eq:complexacvs}
\end{align}
where $z^\ast_t$ denotes the complex conjugate of $z_t$. The sequences $s_{zz}(\tau)$ and $r_{zz}(\tau)$ are in general complex-valued and have the properties that $s_{zz}(\tau)=s^\ast_{zz}(-\tau)$ (Hermitian symmetry) and $r_{zz}(\tau)=r_{zz}(-\tau)$ (symmetry). The sequence $r_{zz}(\tau)$ is sometimes also referred to as the relation sequence \cite{rubin2008kinematics}, or the pseudo-covariance \cite{coeurjolly2016fast}. Complex-valued processes that have the property that $r_{zz}(\tau)=0$ for all $\tau$ are said to be {\em proper}, and are otherwise said to be {\em improper} \cite{schreier2010statistical}. In this paper we focus on the improper case where $r_{zz}(\tau)\neq0$.

The simplest way to relate the sequences in~\eqref{eq:bivariateacvs} and~\eqref{eq:complexacvs} is to start in the complex representation and expand  such that
\begin{align}
s_{zz}(\tau)  &= \E\left\{(x_{t+\tau}+\ri y_{t+\tau})(x_t-\ri y_t)\right\}
\nonumber\\
&=\E\{x_{t+\tau}x_t\}+\E\{y_{t+\tau}y_t\}
\nonumber\\
&\quad\quad+\ri\left(\E\{x_ty_{t+\tau}\}-\E\{x_{t+\tau}y_t\}\right)
\nonumber\\
&=s_{xx}(\tau)+s_{yy}(\tau)+\ri\left[s_{xy}(-\tau)-s_{xy}(\tau)\right]
\label{eq:biv2com1}
\\
r_{zz}(\tau)  &= \E\left\{(x_{t+\tau}+\ri y_{t+\tau})(x_t+\ri y_t)\right\}
\nonumber\\
&=E\{x_{t+\tau}x_t\}-\E\{y_{t+\tau}y_t\}
\nonumber\\
&\quad\quad+\ri\left(\E\{x_ty_{t+\tau}\}+\E\{x_{t+\tau}y_t\}\right)
\nonumber\\
&=s_{xx}(\tau)-s_{yy}(\tau)+\ri\left[s_{xy}(-\tau)+s_{xy}(\tau)\right]
\label{eq:biv2com2}
\end{align}
where for the complex-valued process to be proper we therefore require $s_{xx}(\tau)=s_{yy}(\tau)$ and $s_{xy}(\tau)=-s_{xy}(-\tau)$.

\section{THE METHOD}
\label{sec:method}
We provide the pseudo-code for the simulation algorithm of complex-valued improper processes in Algorithm~\ref{mainAlgorithm}. Then in Sections~\ref{SS1}--\ref{SS3} we detail each line of the algorithm, followed by discussions in Section~\ref{SS4}. We also provide a corresponding MATLAB routine for simulating improper process which can be downloaded from \texttt{http://ucl.ac.uk} \texttt{/statistics/research/spg/software}.

\begin{algorithm*}
\caption{\label{mainAlgorithm} Pseudo-code for the simulation algorithm for improper processes}
\begin{spacing}{1.2}
\begin{algorithmic}[1]
\STATE {\bf INPUTS:} $s_{zz}(\tau)$, $r_{zz}(\tau)$, for $\tau=0,\ldots,n$
\FOR{$\tau=0$ to $n$}
\STATE $s_{xx}(\tau)\leftarrow \Re\left\{\frac{1}{2}\left[s_{zz}(\tau)+r_{zz}(\tau)\right]\right\}$; $\quad s_{yy}(\tau)\leftarrow \Re\left\{\frac{1}{2}\left[s_{zz}(\tau)-r_{zz}(\tau)\right]\right\}$
\STATE $s_{xy}(\tau)\leftarrow \Im\left\{\frac{1}{2}\left[r_{zz}(\tau)-s_{zz}(\tau)\right]\right\}$; $\quad s_{xy}(-\tau)\leftarrow \Im\left\{\frac{1}{2}\left[s_{zz}(\tau)+r_{zz}(\tau)\right]\right\}$
\ENDFOR
\STATE ${\bf c_{xx}} \leftarrow [s_{xx}(0),s_{xx}(1),\ldots,s_{xx}(n-1),s_{xx}(n),s_{xx}(n-1),\ldots,s_{xx}(1)]$
\STATE ${\bf c_{yy}} \leftarrow [s_{yy}(0),s_{yy}(1),\ldots,s_{yy}(n-1),s_{yy}(n),s_{yy}(n-1),\ldots,s_{yy}(1)]$
\STATE ${\bf c_{xy}} \leftarrow [s_{xy}(0),s_{xy}(-1),\ldots,s_{xy}(-(n-1)),s_{xy}(-n),s_{xy}(n-1),\ldots,s_{xy}(1)]$
\STATE ${\bf {\bm\lambda}_{xx}} \leftarrow \mathrm{FFT}({\bf c_{xx}})$; $\quad{\bf {\bm\lambda}_{yy}} \leftarrow \mathrm{FFT}({\bf c_{yy}})$; $\quad{\bf {\bm\lambda}_{xy}} \leftarrow \mathrm{FFT}({\bf c_{xy}})$
\STATE Generate 4 length-$2n$ i.i.d. Gaussian sequences with mean 0 and variance 1/2, denote as: ${\bf w_1}$, ${\bf w_2}$, ${\bf w_3}$, ${\bf w_4}$
\FOR{$k=1$ to $2n$}
\vspace{2mm}
\STATE $\Sigma_k\leftarrow2\begin{pmatrix}1 & \lambda_{xy}(k)/{\sqrt{\lambda_{xx}(k)\lambda_{yy}(k)}}\\\lambda^\ast_{xy}(k)/{\sqrt{\lambda_{xx}(k)\lambda_{yy}(k)}} & 1 \end{pmatrix}$; $\quad A_k\leftarrow\texttt{chol}(\Sigma_k)$
\vspace{2mm}
\STATE $\begin{bmatrix}g_x(k)\\g_y(k)\end{bmatrix}\leftarrow A_k\begin{bmatrix}w_1(k)+\ri w_2(k)\\w_3(k)+\ri w_4(k)\end{bmatrix}$; $\quad\begin{bmatrix}h_x(k)\\h_y(k)\end{bmatrix}\leftarrow\frac{1}{\sqrt{2n}}\begin{bmatrix}\lambda_{xx}(k)g_x(k)\\\lambda_{yy}(k)g_y(k)\end{bmatrix}$
\vspace{2mm}
\ENDFOR
\STATE ${\bf q_x}\leftarrow \mathrm{FFT}({\bf h_{x}})$; $\quad{\bf q_y}\leftarrow \mathrm{FFT}({\bf h_{y}})$
\STATE
${\bf z_1}\leftarrow[\Re\{q_x(1)\},\ldots,\Re\{q_x(n)\}]+\ri [\Re\{q_y(1)\},\ldots,\Re\{q_y(n)\}]$
\STATE
${\bf z_2}\leftarrow[\Im\{q_x(1)\},\ldots,\Im\{q_x(n)\}]+\ri [\Im\{q_y(1)\},\ldots,\Im\{q_y(n)\}]$
\STATE {\bf RETURN:} ${\bf z_1}$, ${\bf z_2}$
\end{algorithmic}
\end{spacing}
\end{algorithm*}

\subsection{Converting from complex to bivariate}\label{SS1}
To generate a complex-valued sequence of length $n$ of a zero-mean Gaussian process we first specify $s_{zz}(\tau)$ and $r_{zz}(\tau)$ for $\tau=0,\ldots,n$ (in Algorithm~\ref{mainAlgorithm}, line 1). This fully specifies the properties of the generated sequence as negative lags up to $-n$ are obtained through the properties $s_{zz}(\tau)=s^\ast_{zz}(-\tau)$ and $r_{zz}(\tau)=r_{zz}(-\tau)$, and lags larger than $\pm n$ are not required.
The next step of our approach (in Algorithm~\ref{mainAlgorithm}, lines 2--5) is to specify the complex-valued process $z_t$ in terms of the autocovariances and cross-covariances of $x_t$ and $y_t$, which can be performed by inverting the relationships in~\eqref{eq:biv2com1} and~\eqref{eq:biv2com2} yielding
\begin{align}
s_{xx}(\tau)&= \Re\left\{\frac{1}{2}\left[s_{zz}(\tau)+r_{zz}(\tau)\right]\right\} \label{eq:com2biv1}\\
s_{yy}(\tau)&= \Re\left\{\frac{1}{2}\left[s_{zz}(\tau)-r_{zz}(\tau)\right]\right\} \\
s_{xy}(\tau)&= \Im\left\{\frac{1}{2}\left[r_{zz}(\tau)-s_{zz}(\tau)\right]\right\} \\
s_{xy}(-\tau)&= \Im\left\{\frac{1}{2}\left[s_{zz}(\tau)+r_{zz}(\tau)\right]\right\}, \label{eq:com2biv4}
\end{align}
for $\tau = 0,\ldots,n$, where $\Re\{\cdot\}$ and $\Im\{\cdot\}$ denote the real and imaginary part respectively. Note that $s_{xx}(\tau)$ and $s_{yy}(\tau)$ are defined for $\tau=-n,\ldots,-1$ by symmetry, i.e. $s_{xx}(\tau)=s_{xx}(-\tau)$ and $s_{yy}(\tau)=s_{yy}(-\tau)$.

\subsection{Generating the bivariate series}\label{SS2}
Having specified $s_{xx}(\tau)$, $s_{yy}(\tau)$, and $s_{xy}(\tau)$, for $\tau=-n,\ldots,0,\ldots,n$, we proceed to use the method of \cite{helgason2011fast} for multivariate Gaussian time series, setting the dimension parameter to $p=2$ for bivariate series. We detail this within Algorithm~\ref{mainAlgorithm}, where we have simplified the algorithm provided in \cite{helgason2011fast} for multivariate processes to bivariate processes. 

In Algorithm~\ref{mainAlgorithm}, lines 6--8, the length-2$n$ vectors $\bf{c_{xx}}$, $\bf{c_{xy}}$, and $\bf{c_{xy}}$, are the top rows of the embedded circulant matrices, where the full matrices do not need to be formulated (thus reducing storage requirements from 4$n^2$ to 2$n$). The Fast Fourier Transform (FFT) of these vectors diagonalizes the circulant matrices, yielding the respective sequences of eigenvalues $\bf{\bm\lambda_{xx}}$, $\bf{\bm\lambda_{yy}}$, and $\bf{\bm\lambda_{xy}}$, which are computed in line 9 of Algorithm~\ref{mainAlgorithm} via the operation
\begin{equation}
\lambda_{xx}(k)=\sum_{j=0}^{2n-1}c_{xx}(k)\exp\{-\ri 2\pi j(k-1)/2n\},
\label{eq:fft}
\end{equation}
for $k=1,\ldots,2n$, and similarly for $\lambda_{yy}(k)$ and $\lambda_{xy}(k)$. Note that $\lambda_{xx}(k)$ and $\lambda_{yy}(k)$ are real-valued, whereas $\lambda_{xy}(k)$ is in general complex-valued. As each FFT is of length $2n$, the ``powers of two" FFT algorithm can be used when $n$ is set to be a power of two.

The next step (Algorithm~\ref{mainAlgorithm}, line 10) is to generate four length-$2n$ i.i.d Gaussian sequences with mean 0 and variance 0.5, which we denote as $\bf{w_1}$, $\bf{w_2}$, $\bf{w_3}$, and $\bf{w_4}$. Then a length-$2n$ for loop is performed (Algorithm~\ref{mainAlgorithm}, line 11--14) to correlate the random variables correctly. First, in line 12, the $2\times2$ matrix $\Sigma_k$ is computed using the eigenvalue sequences. A remark made in \cite{helgason2011fast} is that if $\lambda_{xx}(k)=0$ or $\lambda_{yy}(k)=0$, then the off-diagonals in $\Sigma_k$ should be set to 0, and this does not lose the exactness of the procedure. The next step is to compute the Cholesky decomposition, denoted as $A_k=\texttt{chol}(\Sigma_k)$. Then in Line 13, the matrix $A_k$ is used to transform the random Gaussian sequences, $\bf{w_1}$, $\bf{w_2}$, $\bf{w_3}$, and $\bf{w_4}$, into two correlated complex-valued length-$2n$  vectors $\bf{g_x}$ and $\bf{g_y}$. Finally, these are normalized by the eigenvalue sequences, $\bf{\bm\lambda_{xx}}$ and $\bf{\bm\lambda_{yy}}$, yielding the two vectors $\bf{h_x}$ and $\bf{h_y}$.

The final step (Algorithm~\ref{mainAlgorithm}, line 15) is to transform back from the frequency domain to the time domain, by performing the FFT (as in~\eqref{eq:fft}) on $\bf{h_x}$ and $\bf{h_y}$. This yields two length-$2n$ vectors ${\bf q_x}$ and ${\bf q_y}$.

\subsection{Generating the complex-valued series}\label{SS3}
To generate a complex-valued sequence of length $n$, with prescribed autocovariance $s_{zz}(\tau)$ and complementary covariance $r_{zz}(\tau)$, we take the real parts of the first $n$ values of ${\bf q_x}$ and ${\bf q_y}$, to form a complex-valued sequence ${\bf z_1}$, as given in Algorithm~\ref{mainAlgorithm}, line 16. In addition, we can recover a second sequence ${\bf z_2}$ by taking the first $n$ values of the {\em imaginary} parts of ${\bf q_x}$ and ${\bf q_y}$ (Algorithm~\ref{mainAlgorithm}, line 17). Therefore to generate $M$ length $n$ sequences with the same $s_{zz}(\tau)$ and $r_{zz}(\tau)$, we only need to run the algorithm $\lceil M/2 \rceil$ times.

\subsection{Discussion}\label{SS4}
The exactness of the approach of Algorithm~\ref{mainAlgorithm} follows directly from the exactness of the method proposed in \cite{helgason2011fast}, together with the derivation of~\eqref{eq:biv2com1} and~\eqref{eq:biv2com2}. Algorithm~\ref{mainAlgorithm} is executable in $\mathcal{O}(n\log_2 n)$ operations, as all operations can be computed in $\mathcal{O}(n)$, except for the 5 Fourier transforms in Lines 9 and 15, which can be each computed in $\mathcal{O}(n\log_2 n)$ using the standard ``powers of two" FFT algorithm.

As discussed in \cite{percival2006exact,helgason2011fast} and elsewhere, a potentially problematic aspect of implementing circulant embedding algorithms is that the eigenvalues $\bf{\bm\lambda_{xx}}$ and $\bf{\bm\lambda_{yy}}$ (in Algorithm~\ref{mainAlgorithm}, line 9) are not guaranteed to be nonnegative definite. Negative eigenvalues occur when the circulant matrices (in Algorithm~\ref{mainAlgorithm}, line 6--7) are not nonnegative definite, and are hence invalid covariance matrices. Two ways of dealing with negative eigenvalues are commonly suggested in the literature. The first is to over-sample and implement Algorithm~\ref{mainAlgorithm} with $m>n$, where certain values of $m$ may yield entirely nonnegative eigenvalues. The second is to set negative eigenvalues to zero, which renders the methods inexact. 

Note that $\bf{\bm\lambda_{xy}}$ is complex-valued and nonnegative eigenvalues are not required here for the exactness of the procedure. We do however require that the matrix $\Sigma_k$ is nonnegative definite for the Cholesky decomposition to be valid. Here, \cite{helgason2011fast} suggest approximating $\Sigma_k$ with a nonnegative definite matrix found by setting negative eigenvalues of $\Sigma_k$ to zero.

In \cite{helgason2011fast}, the authors prove two scenarios in which the method is guaranteed to be exact, which we briefly discuss here. First we define the $2\times2$ covariance matrix
\[
R_\tau = \begin{pmatrix}s_{xx}(\tau) & s_{xy}(\tau) \\ s_{xy}(-\tau) & s_{yy}(\tau)\end{pmatrix}
\vspace{-1mm}
\]
for all $\tau\in\mathbb{Z}$. The first scenario is when the matrices $R_\tau$ are absolutely summable, and the corresponding spectral density matrix is positive definite. Then the method is exact for large enough $m$. The second scenario is in time-reversible cases, which in the bivariate case simply means that $s_{xy}(\tau)$ = $s_{xy}(-\tau)$, then the method is exact for any $m\ge n$ if the matrices $R_\tau$, $\Delta R_\tau=R_\tau-R_{\tau+1}$, and $\Delta^2R_\tau=R_\tau-2R_{\tau+1}+R_{\tau+2}$, are nonnegative definite for $\tau\ge0$.

At first glance these conditions appear hard to interpret, but intuition can be developed by looking in the complex domain. By examining~\eqref{eq:com2biv1}--\eqref{eq:com2biv4} we can see that time-reversibility follows when $\Im\{s_{zz}(\tau)\}=0$ for all $\tau$. Then if we also have that $\Im\{r_{zz}(\tau)\}=0$, it follows that $R_\tau$ is diagonal, with entries equal to $s_{zz}(\tau)+r_{zz}(\tau)$ and $s_{zz}(\tau)-r_{zz}(\tau)$. Thus if we meet the three conditions that
\begin{equation}|r_{zz}(\tau)|\le s_{zz}(\tau)
\label{eq:cond1}
\end{equation}
\begin{equation}
|r_{zz}(\tau)-r_{zz}(\tau+1)|\le s_{zz}(\tau)-s_{zz}(\tau+1)
\end{equation}
\vspace{-7mm}
\begin{multline}
|r_{zz}(\tau)-2r_{zz}(\tau+1)+r_{zz}(\tau+2)|\le \\ s_{zz}(\tau)-2s_{zz}(\tau+1)+s_{zz}(\tau+2)
\label{eq:cond3}
\end{multline}
for $\tau\ge0$, then the simulation method is exact. We will revisit these conditions in our applications section.

\section{Application to improper fractional Gaussian noise}
In this section we test our algorithm on an improper fractional Gaussian noise (fGn) process. The complex-valued fGn process \cite{mandelbrot1968fractional} is a stationary Gaussian process which, when regularly sampled at timesteps $t\in\mathbb{Z}$, has autocovariance
\begin{equation}
s_{zz}(\tau)=\frac{V_H}{2}A^2\left(|\tau+1|^{2H}+|\tau-1|^{2H}-2|\tau|^{2H}\right)
\label{eq:fgn}
\end{equation}
where $0<H<1$ is known as the Hurst parameter and controls the roughness or smoothness of the process, with a higher value leading to a smoother process. The normalizing parameter $V_H$ is set as
\begin{equation}
V_H=\frac{\Gamma(H)\Gamma(1-H)}{\pi\Gamma(2H+1)}
\label{eq:VH}
\end{equation}
which is required to regulate the behavior of fractional Brownian motion (fBm) \cite{lilly2016fractional}, where fGn is defined as the increment process of a regularly sampled fBm.

Setting $H=0.5$ recovers a regular Gaussian white noise process. When $H>0.5$, then $s_{zz}(\tau)>0$ and the process is said to be {\em persistent}, as it is positively correlated in time. Conversely, if $H<0.5$ then $s_{zz}(\tau)<0$ and the process is {\em anti-persistent} and negatively correlated in time. fGn with $H>0.5$ is a commonly used long memory model in numerous applications, see e.g. \cite{paxson1997fast}.

For our improper process, we propose to model the complementary covariance to follow the same form of $s_{zz}(\tau)$ such that
\begin{equation}
r_{zz}(\tau)=\frac{V_H}{2}B^2\left(|\tau+1|^{2H}+|\tau-1|^{2H}-2|\tau|^{2H}\right)
\label{eq:fgn2}
\end{equation}
where $B^2<A^2$. It is easy to show that this simple model is a valid process by either transforming into the frequency domain and verifying that the spectral matrix is nonnegative definite \cite{sykulski2013whittle}, or by transforming to a bivariate representation using~\eqref{eq:com2biv1}--\eqref{eq:com2biv4}.

\begin{figure}
\vspace{-3mm}
\centering
\includegraphics[width=0.46\textwidth]{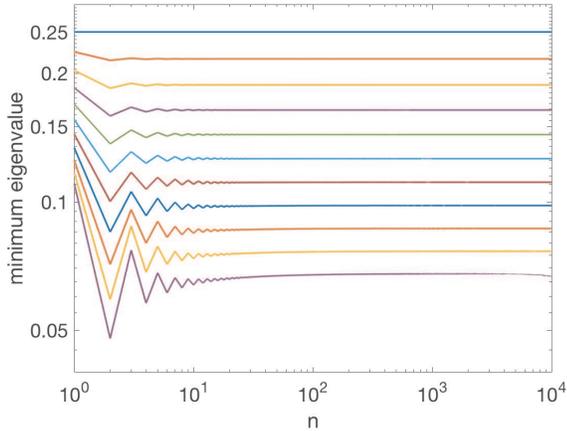}
\vspace{-3mm}
\caption{\label{eigenvalues}The minimum eigenvalue for $\{\bf{\bm\lambda_{xx}},\bf{\bm\lambda_{yy}}\}$ obtained over different values of $n$ in line 9 of Algorithm~\ref{mainAlgorithm} for the models of~\eqref{eq:fgn} and \eqref{eq:fgn2} where $A=1$ and $B=1/\sqrt{2}$. Each line represents a different value of the Hurst parameter $H$, and corresponding normalizing constant $V_H$ in~\eqref{eq:VH}, where the top line is $H=0.5$ and in each subsequent line $H$ increases in increments of 0.05. The bottom line is $H=0.9999$, as the process is not defined for $H=1$, only in the limit as $H\rightarrow1$.}
\vspace{-2mm}
\end{figure}

It also follows that the conditions in~\eqref{eq:cond1}--\eqref{eq:cond3} are satisfied by~\eqref{eq:fgn} and \eqref{eq:fgn2} when $H\ge0.5$. This can be verified by first observing that $s_{zz}(\tau)$ is positive-decreasing for $\tau\ge0$, and is also convex, such that the right hand sides of~\eqref{eq:cond1}--\eqref{eq:cond3} are positive for all $\tau\ge0$. Then as $r_{zz}(\tau)$ is a re-scaled version of $s_{zz}(\tau)$, the terms on the left hand side are always a factor of $B^2/A^2$ smaller than the right hand side, and the three inequalities are hence satisfied. Finally, as $\Im\{s_{zz}(\tau)\}=0$ and $\Im\{r_{zz}(\tau)\}=0$, it follows from the proof of \cite{helgason2011fast} that the circulant embedding procedure of Algorithm~\ref{mainAlgorithm} is guaranteed to be exact for all $n$ with this process. As an additional verification, in Fig.~\ref{eigenvalues} we plot the minimum eigenvalue for $\{\bf{\bm\lambda_{xx}},\bf{\bm\lambda_{yy}}\}$ obtained in Line 9 of Algorithm~\ref{mainAlgorithm} over different values of $H\ge0.5$ and $n$ (with $A=1$ and $B=1/\sqrt{2}$), and they are seen to be positive everywhere.

\begin{figure}
\centering
\includegraphics[width=0.42\textwidth]{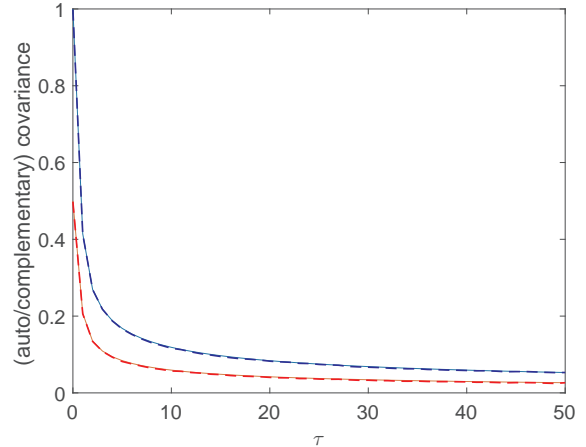}
\vspace{-3mm}
\caption{\label{covariances}The thin solid lines are the autocovariance and cross covariance models of~\eqref{eq:fgn} (in blue) and \eqref{eq:fgn2} (in red) respectively, for the first 50 lags. We have set the parameters to $H=0.75$, $A=1/\sqrt{V_H}$, and $B=A/\sqrt{2}$. Overlaid (in thick dashed lines) are the averaged unbiased auto and complementary covariance estimates for 1000 time series of length $n=1000$, simulated using Algorithm~\ref{mainAlgorithm}.}
\vspace{-2mm}
\end{figure}

\begin{figure}
\centering
\includegraphics[width=0.44\textwidth]{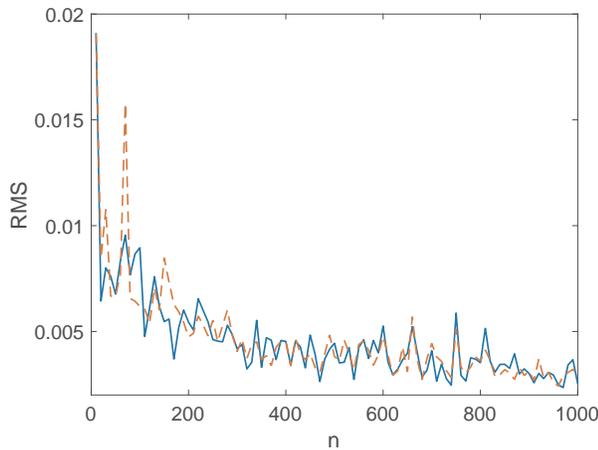}
\vspace{-2.5mm}
\caption{\label{errors}The blue-solid line is the root mean square difference of the theoretical and observed autocovariance, calculated using~\eqref{eq:rms}, averaging over 1000 time series simulated using Algorithm~\ref{mainAlgorithm}, with the same parameter values as in Fig.~\ref{covariances}. The analysis is repeated over values of $n$ ranging from 10 to 1000 in increments of 10. The red-dashed line is the same analysis, this time reporting the root mean square difference of the theoretical and observed complementary covariance.}
\vspace{-2mm}
\end{figure}

We now simulate regularly sampled discrete sequences from the model of~\eqref{eq:fgn} and~\eqref{eq:fgn2}. As the process is not Markovian to any finite order, and hence cannot be represented as an autoregressive process of finite order, then we cannot recursively simulate the sequences in $\mathcal{O}(n)$ operations, and as such we turn to our circulant embedding algorithm. We first simulate 1000 time series each of length $n=1000$. We fix the parameters in~\eqref{eq:fgn} and~\eqref{eq:fgn2} to $H=0.75$, $A=1/\sqrt{V_H}$, and $B=A/\sqrt{2}$, such that the variance is normalized to 1. We display plots of $s_{zz}(\tau)$ and $r_{zz}(\tau)$ in Fig.~\ref{covariances} along with the average unbiased estimators of the autocovariance and complementary covariance for the 1000 simulated time series. The exactness of the method appears to hold, however to test this more robustly we repeat the experiment over values of $n$ ranging from 10 to 1000 (in increments of 10). For each value of $n$, we simulated 1000 time series and report the root mean square difference between $s_{zz}(\tau)$ and its averaged unbiased estimate $\hat{s}_{zz}(\tau;n)$ (as performed in \cite{percival2006exact}) given by
\begin{equation}
RMS(s_{zz}(\tau);n)=\sqrt{\frac{1}{n}\sum_{\tau=0}^{n-1}\left|\hat{s}_{zz}(\tau;n)-s_{zz}(\tau)\right|^2}.
\label{eq:rms}
\end{equation}
We similarly compute $RMS(r_{zz}(\tau);n)$ with each value of $n$. These metrics are displayed in Fig.~\ref{errors}, and are seen to decay as $n$ increases. Even for small $n$ however, the root mean square distance is less than 0.02, supporting the statement that for this process the method is an exact procedure for all $n$.

\section{Conclusions}
\vspace{-2mm}
Improper or noncircular Gaussian processes have been receiving increasing attention in signal processing \cite{adali2011complex}. In this paper, an algorithm for generating improper complex-valued sequences has been proposed for stationary Gaussian processes specified in the time domain. This builds on previous methods for proper processes \cite{percival2006exact,coeurjolly2016fast}, and with improper processes modeled in the frequency domain \cite{chandna2013simulation}. Our method is a reformulation of \cite{helgason2011fast}, where we have simplified the approach to bivariate processes, and then converted into the complex domain. We discussed how theoretical conditions under which exactness is guaranteed extend to the complex representation. An example of such a model was given, by proposing a simple improper complex fractional Gaussian noise process, and simulation evidence supported the exactness of the algorithm.

Circulant embedding simulation methods are only exact when computed eigenvalues from circulant matrices are nonnegative. As reported in \cite{gneiting2012fast} for spatial processes, the issue of negative eigenvalues appears to grow going from one to two dimensions (and indeed higher). As an example, we implemented the improper periodic covariance kernel for complex-valued time series used in \cite{tobar2015modelling} for climate modeling, and found negative eigenvalues were unavoidable with any $n$. Setting these to zero, the root mean square metric of \eqref{eq:rms} was found to be much higher, at around 0.1 for $n=1000$. Therefore, pending more use in practice, the limitations of the simulation algorithm remain unclear. Important avenues of future investigation are to find more general conditions under which exactness does and does not hold, as well as to see if the error bounds of approximate methods can be controlled theoretically when exact $\mathcal{O}(n\log n)$ simulation is not achievable.

{\centering \section*{{\normalsize ACKNOWLEDGMENTS}}}
\vspace{-1mm}
The authors would like to thank Dr Swati Chandna for providing her insightful comments and expertise in this area.
\vspace{5mm}

\bibliographystyle{IEEEbib}

\end{document}